# Playing The Ethics Card: Ethical Aspects In Design Tools For Inspiration And Education

Albrecht Kurze

Chemnitz University of Technology, Albrecht.Kurze@informatik.tu-chemnitz.de

Arne Berger

Anhalt University of Applied Sciences, Arne.Berger@hs-anhalt.de

This paper relates findings of own research in the domain of co-design tools in terms of ethical aspects and their opportunities for inspiration and in HCI education. We overview a number of selected general-purpose HCI/design tools as well as domain specific tools for the Internet of Things. These tools are often card-based, not only suitable for workshops with co-designers but also for internal workshops with students to include these aspects in the built-up of their expertise, sometimes even in a playful way.

CCS CONCEPTS • Human-centered computing~Human computer interaction (HCI);

**Additional Keywords and Phrases:** design; design methods; design tools; design process; education; ideation; Internet of Things; IoT; ethics; UX; privacy.

## 1   INTRODUCTION

The theme of the World Usability Day 2018 was "UX Design for Good or Evil?" and it offered a series of insights, e.g. a report from HCI/UX professionals for 'optimizing' a price comparison website. It is quite apparent that HCI/UX expertise has quite an ambivalent nature of 'Good and Evil', often at the same time. It is sometimes a fine line between supporting users on the one hand side and deceiving and manipulating them on the other. Good and Evil are often not easy to distinguish and may occur at the same time. UX professionals have the knowledge and expertise to cater for both sides. However, facing such ethical challenges is not limited to websites or apps as often discussed in UX design and HCI education. Especially the Internet of Things (IoT) introduces a whole bunch of additional challenges with ubiquitous and connected devices, and sensors that capture data on all aspects of human life. Privacy intrusions and ubiquitous surveillance are only a few of the risks. They also deserve to be included for raising awareness in co-design activities and in future HCI professionals performing them.

## 2   FACING ETHICAL CHALLENGES WHEN DESIGNING AND DEVELOPING FOR THE IOT

In our own participatory and design-led research for the Internet of Things (IoT) in the context of the home, we encountered a plethora of ethical challenges. The materiality of co-design toolkits for the IoT sometimes encouraged co-designers to come up with morally questionable design scenarios, once they realized the possibilities of IoT technology in their home life. We found in our participatory design approaches numerous examples. In ideating situated IoT design scenarios for the home, co-designers came up with the 'automated rent debtor', giving people cold showers when their rent is overdue [6]. Likewise, in sharing simple sensor data in communal living, co-designers started to use sensor technology to nudge their housemates to separate their waste correctly [10]. Other co-designers in our inquiries used simple sensor data to nudge family members to save energy and even used sensor data to monitor family members on when they left the home or came back [17]. Often, such design scenarios go unnoticed, because design researchers do not report ethically questionable design scenarios and report situated and heterogeneous design scenarios instead [6].

## 3 USING DESIGN METHODS AND TOOLS FOR CO-DESIGN AND EDUCATION

A number of design methods and tools exist that makes basic principles and patterns of UX design usable – sometimes even in a fun way, e.g. in the format of card decks or even games. We have collected and reflected on a number of such methods and tools in the last couple of years, mainly with a focus on designing for the IoT. To that end we conducted a number of participatory workshops with design researchers, to explore if they are fit IoT co-design [5,18]. In revisiting this collection, we now want to focus on whether and how ethical issues are addressed, either explicitly or implicitly.

The methods and tools might help addressing ethical aspects on different levels…
1) from (ethical) core values
2) over principles
3) up to best practices

… and in different ways …
a) as a collection of guidelines or hints
b) helping to ask the right questions for inspiration or reflection

… as well as with different purposes …
i) helping to become aware of potential problems and implications otherwise overseen or forgotten
ii) helping to decide for the good or even better (in sense of ethics)

Following on this inquiry, we curated a selection of some such methods and tools in order to give an overview of how these methods and tools help to consider ethical issues. These tools are often card-based and suitable both for workshops with co-designers, as well as students, and are helpful to building up expertise in IoT design, often in a playful way.

## 4 EXAMPLES OF GENERAL PURPOSE DESIGN METHODS AND TOOLS FOR HCI AND UX

Focusing on ethics, some domain unspecific methods and tools for HCI/UX design exist and they help to some extent in reflecting psychological, social, and ethical considerations of technology design.

*Mental Notes* [2] is a tool focusing on psychological aspects in UX. The 54 cards provide background for usable UX principles and examples in different basic categories (attention, comprehension, memory, persuasion, and understanding). Most of the principles seem to serve the evil side (e.g. how to persuade a user to buy more etc.). Nevertheless, the cards can be useful to raise awareness for potentially problematic and ambivalent UX practices and can be a viable source of inspiration for reverse thinking (e.g. to avoid an unwanted persuasion of a user). **Behavior Change Strategy Cards** [4] and **Design with Intent toolkit** [20] have very similar concepts like Mental Notes. **Behavior Change Design Sprints** [7,8] can frame these tools methodically and explicitly introduces ethics for the evaluation of certain strategies (e.g. persuasion) and (dark) UX patterns.

**Moral Agent** "is an ethics ideation game based on brainwriting and hidden roles" [14] and part of '*Ethics for Designers Toolkit*' [15]. The cards introduce 20 moral values (e.g. equity or privacy) and help to ask important question in a very simple scheme of 'How might the design…" e.g. "…allow the users to adjust their privacy settings?". In the game each player is an ambassador of a moral value on a randomly drawn card and has to bring in the moral value into a design while identifying the hidden moral values of other players.

**Tarot Cards of Tech** [3] foster changes of perspectives in design by providing 8 role cards (e.g. as a 'big bad wolf' or a 'backstabber') to reveal unintended consequences of technology. The cards ask tricky questions that directly address potential implications, e.g. "Who loses their jobs?" or "What could make people feel unsafe or exposed?" [3]. Using this reverse thinking can help to "reveal opportunities for creating positive change" [3].

**Guidelines for Human-AI-Interaction** [1,22] should help to brainstorm and evaluate ideas involving AI in HCI and UX design. The guidelines are presented in the format of 18 principles and best-practice examples on cards in 4 categories, e.g. "Make clear why the system did what it did."

**Privacy Ideation Cards** [21,28] not only brings the 'system' and 'user' in card decks in play but also uses categories for 'constraints' and 'regulation' in design. The cards give primarily inspiration, e.g. to design for otherwise marginalized users groups (e.g. elderly people, women, childs, etc.), and ask questions even on complex topics such as 'third country data transfer', 'explicit consent' or 'data anonymization'.

**Design Mindset Cards** [16] structure the design process in 6 stages and help to ask the right questions for a good design in the beginning of the design process ('discovery' stage) as well in the end ('reflection' stage).

**Actionable Futures Toolkit** [26] provides a set of canvases for 10 methods in 3 design phases. The toolkit uses reverse thinking to help understand unwanted future scenarios ('anti-futures') and to turn them into desirable futures.

## 5 EXAMPLES OF IOT DOMAIN SPECIFIC DESIGN METHODS AND TOOLS

The methods and tools we present in this section are specific to the IoT design space and only some of them involve reflective aspects on ethics, values and assumptions. None of these methods and tools is complete in any aspect of the design process [19] but they have the potential to complement each other and to be complemented by the methods and tools presented in section 4.

**Mapping the IoT** [29,30] supports the design process by asking the designer questions, e.g. about the product, its context and users on two levels. A first level ensures a basic consideration of these aspects, e.g. for 'Data Value': "How users will benefit from the data produced…?". A second level of "What if" deepens the consideration in these aspects, e.g. "Will data be accessible from third parties?".

**IoT Design Deck** [12,13] includes a set of 'Threats' cards, e.g. for technical challenges but also for ethical concerns (e.g. 'privacy' or 'conflict of needs'). Each of these cards gives an explanation and justifies the importance of consideration in the design process giving examples or questions for reflection. The cards are intended to reveal otherwise hidden problems and to raise awareness for potentially implications.

**IoT Design Kit** [9,27] uses a set of 'wildcards' to let designer reflect on ideas and solutions they developed. Similar to the 'Threats' cards the *IoT Design Deck* these cards either address technical issues or non-technical aspects asking critical questions (e.g. legal aspects "A new law enforces that no personal data can be stored" or aspects of sustainability "Is it possible to […] repair […]?").

*Tiles IoT Toolkit* [23–25] uses a small set of 'criteria' cards asking a main question for reflection (e.g. sustainability) while giving secondary question on how to judge in detail (e.g. how a product lifecycle looks like).

*Better Things / Gold Standard* [31,32] is a framework with a set of 50 cards offering reflection questions in 9 categories (e.g. privacy or sustainability) that help to address important for IoT products, e.g. "Can you stop the device from recording data?".

*Better IoT* [11] is a set of 33 principles in 8 basic categories representing different values (e.g. privacy, security, transparency). *Better IoT* differentiates the principles in three level of 'must have' (e.g. all privacy), 'nice to have', and 'best case' (e.g. in openness).

## 6   CONCLUSION

The methods and tools presented above can inspire how to address ethical aspects in (co-)design processes - for professionals as well as in education. These tools and methods allow introducing these aspects in a usable and often playful way. They help to learn and understand basic principles for critical reflection in terms of ethics and they stir users of these tools and methods in "the right" direction for asking ethics and value-minded questions.

While the IoT specific tools mostly cover technical, networked and interconnected aspects quite well, their use might benefit from a combination with the more general-purpose tools to enable a deeper level of reflection. There might be good combinations of certain tools, e.g. to cover or detail aspects of privacy otherwise underrepresented.

However, just knowing the names of the tools or having a list of them is not enough to make them effective. Exploring and practicing the use of the different tools, understanding their weaknesses and strengths, will eventually help build up the necessary expertise to select the right tool for a specific purpose in the design process, to make them insightful, beneficial and productive. Therefore, we should address and use these methods and tools in internal workshops with our students and thus provide them with the necessary knowledge to use the methods and tools later on in their research and work.

## 7   OUR EXPECTATIONS FOR THE WORKSHOP

In the workshop we hope for inspiration and discussion:
   a)   how the different tools might be supportive in practical use
   b)   how the tools might be reflected back to ethics theory and more classical approaches
   c)   about other participants' experiences with these tools

### ACKNOWLEDGMENTS

This research is partly funded by the German Ministry of Education and Research (BMBF) under grant number FKZ 16SV7116.

**REFERENCES**

[1] Saleema Amershi, Kori Inkpen, Jaime Teevan, Ruth Kikin-Gil, Eric Horvitz, Dan Weld, Mihaela Vorvoreanu, Adam Fourney, Besmira Nushi, Penny Collisson, Jina Suh, Shamsi Iqbal, and Paul N. Bennett. 2019. Guidelines for Human-AI Interaction. In *Proceedings of the 2019 CHI Conference on Human Factors in Computing Systems - CHI '19*, 1–13. https://doi.org/10.1145/3290605.3300233

[2] Stephen P. Anderson. 2019. Mental Notes. Retrieved February 11, 2019 from http://getmentalnotes.com/

[3] Artefact. 2017. The Tarot Cards Of Tech. *Artefact | Product, UX, Industrial Design + Strategy*. Retrieved May 31, 2019 from http://tarotcardsoftech.artefactgroup.com/

[4] Artefact. Behavior change strategy cards. *Artefact*. Retrieved February 21, 2020 from https://www.artefactgroup.com/case-studies/behavior-change-strategy-cards/

[5] Arne Berger and Albrecht Kurze. 2019. How do we ideate responsible IoT products? *ThingsCon*. Retrieved February 20, 2020 from https://thingscon.org/track-b-how-do-we-ideate-responsible-iot-products/

[6] Arne Berger, William Odom, Michael Storz, Andreas Bischof, Albrecht Kurze, and Eva Hornecker. 2019. The Inflatable Cat: Idiosyncratic Ideation Of Smart Objects For The Home. In *CHI Conference on Human Factors in Computing Systems Proceedings*. https://doi.org/10.1145/3290605.3300631

[7] Lucas Colusso. 2019. Behavior Change Design Sprints. *Medium*. Retrieved February 21, 2020 from https://medium.com/ixda/behavior-change-design-sprints-d5f130275b52

[8] Lucas Colusso, Tien Do, and Gary Hsieh. 2018. Behavior Change Design Sprints. In *Proceedings of the 2018 on Designing Interactive Systems Conference 2018 - DIS '18*, 791–803. https://doi.org/10.1145/3196709.3196739

[9] Dries De Roeck, Jürgen Tanghe, Alexis Jacoby, Ingrid Moons, and Karin Slegers. 2019. Ideas of Things: The IOT Design Kit. In *Proceedings of the 2019 Designing Interactive Systems Conference*, 159–163. https://doi.org/10.1145/3301019.3323888

[10] Teresa Denefleh. 2019. Sensorstation. In *Proceedings of Mensch Und Computer 2019* (MuC'19), 871–873. https://doi.org/10.1145/3340764.3345370

[11] Alexandra Deschamps-Sonsino. 2018. Better IoT. *Better IoT*. Retrieved May 31, 2019 from https://betteriot.wordpress.com/

[12] Massimiliano Dibitonto, Katarzyna Leszczynska, and Federica Tazzi. 2019. The IoT Design Deck – A Co-Design method for the connected products. Retrieved February 6, 2019 from http://www.iotdesigndeck.com/

[13] Massimiliano Dibitonto, Federica Tazzi, Katarzyna Leszczynska, and Carlo M. Medaglia. 2017. The IoT Design Deck: A Tool for the Co-design of Connected Products. In *Advances in Usability and User Experience* (Advances in Intelligent Systems and Computing), 217–227. https://doi.org/10.1007/978-3-319-60492-3_21

[14] Jet Gispen. 2017. Moral Agent. *Ethics for Designers*. Retrieved May 31, 2019 from https://www.ethicsfordesigners.com/tools-1/moral-agent

[15] Jet Gispen. 2017. Ethics for Designers. *Ethics for Designers*. Retrieved February 20, 2020 from https://www.ethicsfordesigners.com

[16] InGlobal Learning Design. 2020. Design Mindset CARDS. *InGlobal Learning Design*. Retrieved February 20, 2020 from http://inglobal.org/design-mindset-cards/

[17] Albrecht Kurze, Sören Totzauer, Michael Storz, Maximilian Eibl, Margot Brereton, and Arne Berger. 2020. Guess The Data: Data Work To Understand How People Make Sense Of And Use Simple Sensor Data From Homes. In *Proceedings of the 2020 CHI Conference on Human Factors in Computing Systems* (CHI'20). https://doi.org/10.1145/3313831.3376273

[18] Albrecht Kurze and Sören Totzauer. 2018. IoT Ideation Expert Workshop. Retrieved February 20, 2020 from https://nebeneinander-miteinander.de/iot-ideation-workshop-2018/

[19] Albrecht Kurze, Sören Totzauer, Alexandra Deschamps-Sonsino, and Arne Berger. 2019. A Collaborative Landscaping Exercise of IoT Design Methods. In *Proceedings of the 2019 OZCHI Australian Conference on Human Computer Interaction*. https://doi.org/10.1145/3369457.3369484

[20] Dan Lockton, David Harrison, and Neville A. Stanton. 2010. Introduction to the Design with Intent toolkit. *Design with Intent*. Retrieved February 21, 2020 from http://designwithintent.co.uk/introduction-to-the-design-with-intent-toolkit/

[21] Ewa Luger, Lachlan Urquhart, Tom Rodden, and Michael Golembewski. 2015. Playing the Legal Card: Using Ideation Cards to Raise Data Protection Issues within the Design Process. In *Proceedings of the 33rd Annual ACM Conference on Human Factors in Computing Systems - CHI '15*, 457–466. https://doi.org/10.1145/2702123.2702142

[22] Microsoft. Guidelines for Human-AI Interaction. *Microsoft Research*. Retrieved February 21, 2020 from https://www.microsoft.com/en-us/research/project/guidelines-for-human-ai-interaction/

[23] Simone Mora, Monica Divitini, and Francesco Gianni. 2016. Tiles: An Inventor Toolkit for Interactive Objects. In *Proceedings of the International Working Conference on Advanced Visual Interfaces* (AVI '16), 332–333. https://doi.org/10.1145/2909132.2926079

[24] Simone Mora, Francesco Gianni, and Monica Divitini. 2017. Tiles: A Card-based Ideation Toolkit for the Internet of Things. In *Proceedings of the 2017 Conference on Designing Interactive Systems* (DIS '17), 587–598. https://doi.org/10.1145/3064663.3064699

[25] Simone Mora, Francesco Gianni, and Monica Divitini. 2019. Tiles IoT Inventor Toolkit. Retrieved February 6, 2019 from http://tilestoolkit.io/

[26] Nordkapp. Actionable Futures Toolkit. *The Actionable Futures Toolkit*. Retrieved February 21, 2020 from https://futures.nordkapp.fi/

[27] Studio Dott. 2019. IoT Design Kit. *Studio Dott*. Retrieved April 25, 2019 from https://iotdesignkit.studiodott.be/

[28] Lachlan Urquhart. Information Privacy by Design Cards. *The University of Nottingham*. Retrieved February 20, 2020 from

    https://www.nottingham.ac.uk/research/groups/mixedrealitylab/projects/information-privacy-by-design-cards.aspx
[29] Ilaria Vitali, Venanzio Arquilla, and Valentina Rognoli. 2019. Mapping the IoT toolkit. *Mapping The IoT Toolkit*. Retrieved February 6, 2019 from http://mappingtheiot.polimi.it/
[30] Ilaria Vitali, Valentina Rognoli, and Venanzio Arquilla. 2016. Mapping the IoT: Co-design, Test and Refine a Design Framework for IoT Products. In *Proceedings of the 9th Nordic Conference on Human-Computer Interaction* (NordiCHI '16), 142:1-142:3. https://doi.org/10.1145/2971485.2987681
[31] Noam Zomerfeld. 2016. Gold Standard. Retrieved February 21, 2020 from http://things.zomerfeld.org/
[32] Noam Zomerfeld. 2017. Better Things - Process Documentation. Retrieved August 11, 2018 from http://sites.zomerfeld.org/767580/